\documentclass[]{interact}

\usepackage{epstopdf}
\usepackage[caption=false]{subfig}

\usepackage[numbers,sort&compress]{natbib}
\bibpunct[, ]{[}{]}{,}{n}{,}{,}
\renewcommand\bibfont{\fontsize{10}{12}\selectfont}

\usepackage[autostyle, english = american]{csquotes}
\usepackage{placeins}
\usepackage{xcolor}
\usepackage{mathtools}

\MakeOuterQuote{"}

\DeclarePairedDelimiter\paren{(}{)}

\newcommand{\E}{\mbox{\bf E}}

\theoremstyle{plain}

\theoremstyle{definition}

\theoremstyle{remark}
\newtheorem{remark}{Remark}

\begin{document}


\title{Exact Inference for Disease Prevalence Based on a Test with Unknown Specificity and Sensitivity}

\author{
\name{Bryan Cai\textsuperscript{a}\thanks{CONTACT Bryan Cai. Email: bxcai@stanford.edu}, John P.A. Ioannidis\textsuperscript{b}, Eran Bendavid\textsuperscript{c}, Lu Tian\textsuperscript{d}}
\affil{\textsuperscript{a}Department of Computer Science, Stanford University; \textsuperscript{b,c}Department of Medicine,  Stanford University; \textsuperscript{d}Department of Biomedical Data Science, Stanford University}
}

\maketitle

\begin{abstract}
To make informative public policy decisions in battling the ongoing COVID-19 pandemic, it is important to know the disease prevalence in a population. There are two intertwined difficulties in estimating this prevalence based on testing results from a group of subjects. First, the test is prone to measurement error with unknown sensitivity and specificity. Second, the prevalence tends to be low at the initial stage of the pandemic and we may not be able to determine if a positive test result is a false positive due to the imperfect specificity of the test. The statistical inference based on large sample approximation or conventional bootstrap may not be sufficiently reliable and yield confidence intervals that do not cover the true prevalence at the nominal level. In this paper, we have proposed a set of 95\% confidence intervals, whose validity is guaranteed and doesn't depend on the sample size in the unweighted setting.  For the weighted setting, the proposed inference is equivalent to a class of hybrid bootstrap methods, whose performance is also more robust to the sample size than those based on asymptotic approximations. The methods are used to reanalyze data from a study investigating the antibody prevalence in Santa Clara county, California, which was the motivating example of this research, in addition to several other seroprevalence studies where authors had tried to correct their estimates for test performance. Extensive simulation studies have been conducted to examine the finite-sample performance of the proposed confidence intervals.
\end{abstract}


\section{Introduction}
Determining the proportion of people who have developed antibodies to SARS-CoV-2 due to prior exposure is an important piece of information for guiding the response measures to COVID-19 in different populations. This proportion is also key to understanding the severity of the disease in terms of estimating the infection fatality rate among those infected; for examples see \cite{deSouzaBC20, Ceylan20, SignorelliSO20, HuSD20}, and for an overview of 82 studies see \cite{Ioannidis2020}.  One effective way of estimating the proportion of people who have developed antibodies is to conduct a survey in the community of interest: sample a subgroup of people from the target population and measure their antibody status using a test kit. The prevalence of SARS-CoV-2 antibodies can oftentimes be low, in the range of 0-2\%, especially at the early stage of viral spread in a population. In these circumstances the simple proportion of the positive results among all conducted tests can be a poor estimate of the true prevalence due to the simple fact that all tests are not perfect \cite{sempos20}. For example, suppose that the test used in the study has a fairly good operational characteristics: sensitivity$=95\%$ and specificity$=99\%$. Then, if the true prevalence rate in the study population is 1\%, based on a simple calculation ,  $1\%\times 95\%+99\%\times (1-99\%)=1.9\%$ of the tests would be positive. The proportion of positive tests in this case is almost double the true prevalence.

Therefore, the proportion of the positive tests may not represent the true prevalence and should be adjusted for the sensitivity and specificity of the deployed test. If the sensitivity and specificity of test is unrealistically assumed to be known, an unbiased estimate of the true prevalence can be obtained easily and its 95\% confidence interval can be constructed \cite{ReiczigelFO10}. However, this is usually not the case, and in practice the reported sensitivity and specificity of the test are often obtained based on limited number of experiments and are random estimates subject to errors themselves. In the absence of a gold reference to estimate sensitivity and specificity against, a Bayesian approach is suggested in \cite{EnoeGJ00}. Therefore, it is important to develop an inference procedure to account for the randomness from study data as well as from reported sensitivity and specificity. When the specificity is close to one and the prevalence is low, the statistical inference based on large sample approximation such as delta-method and naive bootstrap may not be reliable. In this paper, we have proposed an exact inference method for the disease prevalence without requiring any large sample approximation. The method can be useful for statistical inference of the prevalence of antibody to SARS-CoV-2 in a given population.
\section{Method}
\subsection{Exact inference based on random sampling}
Our goal is to construct a reliable confidence interval for disease prevalence in scenarios with true specificity close to 1, where the normality assumption for the reported sensitivity does not necessarily hold. In a typical setting, we observe three separate estimates for the proportion of positive tests $r_0$, sensitivity $p_0$, and specificity $q_0$, denoted by 
$\hat{r},$ $\hat{p}$ and $\hat{q}$, respectively.  Specifically, we assume to observe three independent binomials random variables.
\begin{align*} 
d \sim Bin(D, r_0),\\
m \sim Bin(M, p_0), \\
n \sim Bin(N, q_0), 
\end{align*}
$\hat{r}=d/D$, $\hat{p}=m/M$ and $\hat{q}=n/N,$ where $D$ is the sample size in the current study; $M$ is the number of positive reference samples used to estimate the sensitivity; and $N$ is the number of negative reference samples used to estimate the specificity. Normally, $D$ is big relative to $M$ and $N,$ whose typical values are in the range of tens or hundreds. Based on the relationship
$$r_0=\pi_0p_0+(1-\pi_0)(1-q_0),$$
the true disease prevalence satisfies
$$\pi_0=f(r_0, p_0, q_0), $$
where 
$$ f(r, p, q) := \frac{r + q - 1}{p + q - 1}.$$
A simple point estimator of the prevalence $\pi_0$ adjusted for sensitivity and specificity is thus
$$\hat{\pi} = f(\hat{r}, \hat{p}, \hat{q}).$$
If $\min(D, M, N) \rightarrow \infty $ and $r_0, p_0, q_0 \in [0, 1],$ then the central limit theorem suggests the normal approximation:
\[
\begin{bmatrix}
\sqrt{D}(\hat{r} - r_0) \\
\sqrt{M}(\hat{p} - p_0) \\
\sqrt{N}(\hat{q} - q_0)
\end{bmatrix}
\sim N \paren*{
\begin{bmatrix}
0 \\
0 \\
0
\end{bmatrix}
,
\begin{bmatrix}
r_0(1 - r_0) & 0 & 0 \\
0 & p_0(1 - p_0) & 0 \\
0 & 0 & q_0(1 - q_0)
\end{bmatrix}
}\]
By delta method, $\hat{\pi} - \pi_0$ can be approximated by a mean zero Gaussian distribution  $N(0, \sigma_0^2),$
where
\[\sigma_0^2 = \frac{r_0(1 - r_0)}{D(p_0 + q_0 - 1)^2} + \frac{\pi_0^2p_0(1 - p_0)}{M(p_0 + q_0 - 1)^2} + \frac{(r_0 - p_0)^2q_0(1 - q_0)}{N(p_0 + q_0 - 1)^4}.\]
The variance $\sigma_0^2$ is unknown but can be consistently estimated by 
$$\hat{\sigma}^2 = \frac{\hat{r}(1 - \hat{r})}{D(\hat{p} + \hat{q} - 1)^2} + \frac{\hat{\pi}^2\hat{p}(1 - \hat{p})}{M(\hat{p} + \hat{q} - 1)^2} + \frac{(\hat{r} - \hat{p})^2\hat{q}(1 - \hat{q})}{N(\hat{p} + \hat{q} - 1)^4}.$$
Therefore, a simple 95\% confidence interval for $\pi_0$ can be constructed as 
\begin{equation}
\left[\hat{\pi}-1.96 \hat{\sigma},  \hat{\pi}+1.96\hat{\sigma} \right].
\label{eq:CI}\end{equation}
This confidence interval can be viewed as a product of inverting a Wald test $H_0:  \pi_0=\pi$ based on the test statistic
$$\hat{T}(\pi) = \frac{\hat{\pi} - \pi}{\hat{\sigma}}$$
which approximately follows a standard Gaussian distribution with mean zero and unit variance under the null hypothesis $H_0: \pi_0=\pi.$ Specifically, the $p$-value of the test can be calculated as 
$$p(\pi)=P(|Z|>|\hat{T}(\pi)|)$$
and the 95\% confidence interval based on (\ref{eq:CI}) can be expressed as 
$$\{\pi \mid p(\pi)>0.05\},$$
i.e., all $\pi,$ at which we can not reject the null hypothesis at the 0.05 significance level. Here $Z\sim N(0, 1).$

\begin{remark}
In the extreme case, the confidence interval above may include negative values. To preserve the appropriate range of the prevalence, one may first construct the 95\% confidence interval for $logit(\pi_0)$ as
$$ \left[logit(\hat{\pi})-\frac{1.96\hat{\sigma}}{\hat{\pi}(1-\hat{\pi})}, logit(\hat{\pi})+\frac{1.96\hat{\sigma}}{\hat{\pi}(1-\hat{\pi})}\right]$$
and transform it back vai $expit(\cdot)$ to a confidence interval for $\pi_0.$
\end{remark}

When $q_0$ is close to 1 and/or $r_0$ is close to zero, the null distribution of $\hat{T}$ may not be approximated well by the standard normal and the $p$-value based on asymptotic approximation, i.e. $p(\pi),$ becomes unreliable. 
The bootstrap method has also been used to construct the confidence interval of $\pi_0.$ There are different variations of bootstrap method. The simplest version draws 
\begin{align*}
r_b^*&\sim Bin(D, \hat{r})/D\\
p_b^*&\sim Bin(M, \hat{p})/M\\
q_b^*&\sim Bin(N, \hat{q})/N,
\end{align*}
and calculate $\pi^*_b=f(r_b^*, p_b^*, q_b^*)$ for $b=1, \cdots, B,$ where $B$ is a large integer specified by the user.  Then two ends of the 95\% confidence interval of $\pi_0$ can be constructed as 2.5 and 97.5 percentile of $\{\pi^*_b, b=1,\cdots, B\}$. However, the validity of the bootstrap method also rely on large sample approximation, which may be broken in some settings of interest.

To address this concern, we propose to calculate the exact $p$-value by inverting an exact test, while accounting for nuisance parameters as in \cite{ChanZ99,  FeldmanC98, SenWW09, michael2019exact, gronsbell2020exact} through the following steps:   

\begin{enumerate}
    \item  Construct 99.9\% confidence intervals for $r_0, p_0,$ and $q_0$ denoted by $I_{r}, I_{p}$ and $I_q,$ respectively. The high coverage level of 99.9\% is chosen to ensure that the probability of $(r_0, p_0, q_0)\in I_r\times I_p \times I_q$ is much greater than 95\%. Other options such 99.96\% are viable as well. 
    \item  For a given $\pi$,  define  $\Omega_{\pi}=\{(r, p, q) \mid f(r, p, q)=\pi,  (r, p, q) \in I_{r}\times I_{p}\times I_{q}\}.$
    \item  Select a dense net "spanning" $\Omega_{\pi}:$  $\{(r_k, p_k, q_k)\in \Omega_{\pi}, k=1, \cdots, K\}$ such that for any $(r, p, q) \in \Omega_\pi$, there exists a $k\in \{1, \cdots,  K\}$ such that 
    $|r_{k}-r|+|p_{k}-p|+|q_{k}-q|\le \epsilon$ for a small constant $\epsilon>0.$
    \item For each $(r_k, p_k, q_k)$ from the net, we simulate 
    \begin{align*}
r^*_b \sim Bin(D, r_k)/D \\
p^*_b \sim Bin(M, p_k)/M \\
q^*_b \sim Bin(N, q_k)/N,
\end{align*}
and let
\begin{align*}
\pi^*_b&=f(r^*_b, p^*_b, q^*_b)\\
{\sigma^*_b}^2 &= \frac{r_b^*(1 - r^*_b)}{D(p_b^* + q_b^* - 1)^2} + \frac{{\pi^*_b}^2p^*_b(1 - p^*_b)}{M(p^*_b + q^*_b - 1)^2} + \frac{(r^*_b - p^*_b)^2q^*_b(1 - q^*_b)}{N(p^*_b + q^*_b - 1)^4}
\end{align*}
for $b=1, \cdots, B,$ where $B$ is a large number such as $1,000.$
The distribution of $\hat{T}(r_k, p_k, q_k)$ under the simple null hypothesis $H_0: (r_0, p_0, q_0)=(r_k, p_k, q_k),$ can be approximated by the empirical distribution of $$\{T^*_b(r_k, p_k, q_k), b=1, \cdots, B\},$$ where
$$T^*_b(r_k, p_k, q_k)=\frac{\pi^*_b-\pi}{\sigma^*_b}.$$
Specifically, the exact $p$-value for testing $H_0: (r_0, p_0, q_0)=(r_k, p_k, q_k)$ can be estimated by 
$$\hat{p}(r_k, p_k, q_k)=B^{-1}\sum_{b=1}^B I(|T^*_b(r_k, p_k, q_k)|\ge |\hat{T}(\pi)|),$$
where $I(\cdot)$ is the indicator function.
\item Since $H_0: \pi_0=\pi$ is a composite null hypothesis, the exact $p$-value for testing $H_0: \pi_0=\pi$ can be approximated by
$$\hat{p}(\pi)=\max_{k=1, \cdots, K} \hat{p}(r_k, p_k, q_k).$$
\end{enumerate}
Lastly, the 95\% confidence interval for $\pi_0$ can be constructed as 
$$\{\pi \mid \hat{p}(\pi)\ge 0.05-0.003\},$$
i.e., all $\pi$'s with an exact $p$-value is greater than 0.05-0.003=0.047. The adjustment of 0.003 in the significance level is needed to account for joint coverage level of three confidence intervals $I_r,$ $I_p,$ and $I_q$ at step (1), i.e., $P\left((r_0, p_0, q_0)\in I_r\times I_p\times I_q\right)\ge 1-0.003.$ 
\begin{remark}
If $\Omega_{\pi}=\phi,$ i.e., there is no $(r, p, q)\in I_r\times I_p \times I_q$ such that $f(r, p, q)=\pi$, then let $\hat{p}(\pi)=0$.
\end{remark}
\begin{remark}
The construction of the grid points in $\Omega_{\pi}$ can be completed by first considering all the points
$\{(p_i, q_j)\},$ where $\{p_i\}$ is a set of evenly-spaced points over $I_p$ and $\{q_j\}$ is a set of evenly-spaced points over $I_q$.  Then for each $(p_i, q_j)$, we may solve the equation $f(r, p_i, q_j)=\pi$ in terms of $r$.  If the solution $r_{ij}\in I_r,$ then the triplet $(p_i, q_j, r_{ij})$ is included in dense net.
\end{remark}

The proposed exact confidence intervals always can cover the true prevalence at the desired level regardless of the sample size, if we ignore the Monte-Carlo error in calculating the $p$-value $\hat{p}(r_k, p_k, q_k),$ which can be made arbitrarily small by increasing $B$ in the simulation and considering more dense net spanning the region $\Omega_{\pi}.$  

The computation can be slow, since for each hypothesis test, we need to simulate the null distribution of $\hat{T}(\pi)$ for many triplets $(r_k, p_k, q_k)\in \Omega_{\pi}.$ We can accelerate this computation using hybrid bootstrap \cite{ChuangL00}, where some nuisance parameters are fixed to their point estimates when approximating the distribution of test statistic. For example, we may assume $p_0=\hat{p}$ and only consider pairs $(r, q)$ such that $f(r, \hat{p}, q)=\pi$ to generate the exact $p$-value of $H_0: \pi_0=\pi.$ Specifically, we may let $\{q_j, j=1, \cdots, J\}$ be evenly spaced points over $I_q$ and let 
$r_j=\pi \hat{p}-(1-\pi)(q_j-1)$. In the end 
$$\hat{p}(\pi)=\max_{1\le j\le J, r_j \in I_r}  \hat{p}(r_j, \hat{p}, q_j).$$
Since $p_0$ is fixed at $\hat{p},$  the number of pairs $(r, q)$ to be considered for each fixed level $\pi$ can be much smaller than the number of triplets $(r, p, q)$. The price for gaining the computation speed is sacrificing the ``exact'' coverage level in finite sample due to the fact that the observed point estimator of the nuisance parameters may be quite different from the true parameter. Consequently, the coverage level of the confidence interval based on hybrid bootstrap is not always guaranteed in all settings. However, we expect that it still performs better than the naive confidence interval (\ref{eq:CI}), where all unknown parameters were assumed to be their observed estimates.  It can be viewed as a compromise between the computational intensive exact inference and asymptotic inference based on large sample approximations.

\subsection{Hybrid Bootstrap for Weighted Inference}
When the survey for studying disease prevalence is not conducted using a representative sample, appropriate weighting of the samples is needed to obtain an unbiased estimate of the prevalence. There are two typical settings. (1) There are several strata and the sampling is considered random (and thus representative) within each stratum; In such a case, the strata-specific weighting will be employed.  (2) The sampling represents a population different from the target, but a propensity score can be constructed and individual-specific weighting will be needed. In the following, we will address these two cases separately. 
\subsubsection{Stratum Specific Weighting}
Suppose that the target population consists of $S$ strata with proportions $w_1, \cdots, w_{S-1},$ and $w_S.$   Also suppose that the underlying disease prevalence in each of the $S$ strata is $\pi_1, \cdots, \pi_{S-1},$ and $\pi_S.$ The parameter of interest is the overall disease prevalence in the target population:
$$\pi_{w}=\sum_{s=1}^S w_s\pi_s.$$
Let the number of tests conducted in these strata be $D_1, \cdots, D_{S-1},$ and $D_S,$ respectively.  The number of positive tests in stratum $s$ follows a Poisson distribution:   
$$d_s\sim Pois(D_sr_s),$$
which can be approximated by $N(D_sr_s, D_sr_s),$ 
where $r_s=\pi_sp_0+(1-\pi_s)(1-q_0).$
Consequently, 
$$d_w=\sum_{s=1}^S  \tilde{w}_sd_s \sim N\left(Dr_{w}, \lambda_0 Dr_{w}\right)$$
where $D=\sum_{s=1}^S D_s,$ $\tilde{w}_s=w_s/(D_s/D),$ $r_{w}=\sum_{s=1}^S w_s r_s,$ and
$$\lambda_0=\frac{\sum_{s=1}^S \tilde{w}_sw_sr_s}{r_w}$$
is the variance inflation factor.  Noting that $\pi_{w}=f(r_{w}, p_0, q_0),$ we can estimate $\pi_{w}$ based on $d_w$ by 
$$\hat{\pi}_{w}=f(\hat{r}_{w}, \hat{p}, \hat{q}),$$
where $\hat{r}_{w}=d_w/D.$ Define the test statistic
$$\hat{T}_w(\pi)=\frac{\hat{\pi}_{w}-\pi_{w}}{\hat{\sigma}_{w}},$$
where
$$\hat{\sigma}^2_w = \frac{\hat{\lambda}\hat{r}_w}{D(\hat{p} + \hat{q} - 1)^2} + \frac{\hat{\pi}_{w}^2\hat{p}(1 - \hat{p})}{M(\hat{p} + \hat{q} - 1)^2} + \frac{(\hat{r}_{w} - \hat{p})^2\hat{q}(1 - \hat{q})}{N(\hat{p} + \hat{q} - 1)^4},$$
$$\hat{\lambda}=\hat{r}_w^{-1}\left\{\sum_{s=1}^S \tilde{w}_sw_s\hat{r}_s \right\},$$
and $\hat{r}_s=d_s/D_s.$

To  calculate the exact $p$-value for testing $H_0: \pi_{w}=\pi$, we only need to modify the steps (1), (4) and (5) of the algorithm in section 2.1.   
\begin{enumerate}
    \item[(1)]  Construct 99.9\% confidence intervals for $r_{w}$ denoted by $I_{r}$ assuming $\lambda_0=\hat{\lambda}.$
    \item[(4)] For each $(r_k, p_k, q_k)$ from the net for $\Omega_\pi$, we simulate 
    $$ r^*_{wb} \sim N(Dr_k, \hat{\lambda}Dr_k)/D $$
and let
$\pi^*_{wb}=f(r^*_{wb}, p^*_b, q^*_b)$ and
$$\sigma^{*2}_{wb} = \frac{\hat{\lambda}r_{wb}^*}{D(p_b^* + q_b^* - 1)^2} + \frac{{\pi^*_{wb}}^2p^*_b(1 - p^*_b)}{M(p^*_b + q^*_b - 1)^2} + \frac{(r^*_{wb} - p^*_b)^2q^*_b(1 - q^*_b)}{N(p^*_b + q^*_b - 1)^4}$$
for $b=1, \cdots, B.$ 
The exact $p$-value for testing $H_0: (r_{w}, p_0, q_0)=(r_k, p_k, q_k)$ can be estimated by $$\hat{p}_w(r_k, p_k, q_k)=B^{-1}\sum_{b=1}^B I\left( \left| \frac{\pi^*_{wb}-\pi}{\sigma^*_{wb}}\right |\ge |\hat{T}_w(\pi)|\right).$$
\item[(5)] The exact $p$-value for testing $H_0: \pi_{w}=\pi$ can be approximated by
$$\hat{p}_w(\pi)=\max_{k=1, \cdots, K} \hat{p}_w(r_k, p_k, q_k).$$
\end{enumerate}
The 95\% confidence interval for $\pi_{w}$ thus consists of all values of $\pi$ such that the exact $p$ value for testing $H_0: \pi_{w}=\pi$ is greater than $0.05-0.003.$ This procedure assumes that $\lambda_0=\hat{\lambda}$ is known and thus is essentially a hybrid bootstrap method.  Therefore, the inference result may not be exact in all settings. 

\begin{remark}
This method assumes that $d_s\sim Pois(D_sr_s),$ which is approximately Gaussian $N(D_sr_s, D_sr_s).$ Therefore, we implicitly assume that $r_s$ is small and $D_sr_s$ is reasonably big, e.g., $\ge 10.$  If $D_sr_s \approx 0$ and there are very few strata,  it is not clear how to approximate the distribution of $d_w.$ 
\end{remark}

\subsubsection{Individual Specific Weighting}
With a slight abuse of notation, now suppose that the $i$th subject has a test result, denoted by a Bernoulli random variable $d_i\sim Ber(r_i),$ and a weight $w_i$, where  $r_i=\pi_ip_0+(1-\pi_i)(1-q_0),$ $\pi_i$ is probability that the subject has the antibody, and $\sum\limits_{i = 1}^{D} w_i = D.$ Our goal is again to estimate the weighted prevalence
$$\pi_w = \frac{1}{D}\sum\limits_{i = 1}^{D} w_i\pi_i.$$
Since $d_i \sim Ber(r_i)$, as $D$ grows large,  by central limit theorem, we can approximate
$$\frac{1}{D}\sum\limits_{i = 1}^{D} w_id_i \sim N\left(\frac{1}{D}\sum\limits_{i = 1}^{D}w_ir_i, \frac{1}{D^2}\sum\limits_{i = 1}^{D} w_i^2r_i(1 - r_i)\right)=N\left(r_w, \frac{1}{D^2}\sum\limits_{i = 1}^{D} w_i^2r_i(1 - r_i)\right),$$
under the Lindeberg condition that
$$\lim\limits_{D \to \infty} \frac{1}{s_n^2}\sum\limits_{k = 1}^{D}\E[w_k^2(d_k - r_k)^2 I\left\{w_k|d_k - r_k| > \epsilon s_n\right\}] \to 0$$
for any $\epsilon > 0$, where $r_w = D^{-1}\sum\limits_{i = 1}^{D} w_ir_i$ and $s_n = \sum\limits_{k = 1}^{D} w_k^2r_k(1 - r_k)$. A  sufficient condition for the aforementioned convergence is 
$$\lim_{D\rightarrow +\infty}\frac{\max_{1\le j\le D}w_j}{\sum\limits_{j = 1}^{D} w_j^2 r_j (1 - r_j)} \to 0.$$
Since it is difficult to estimate $\sum\limits_{i = 1}^{D} w_i^2r_i(1 - r_i)$, we make the further approximation
\begin{equation}
 \lambda_0 = \frac{\sum_{i = 1}^{D} w_i^2r_i(1 - r_i)}{Dr_w(1-r_w)} \approx  \frac{\sum_{i = 1}^{D} w_i^2r_i}{Dr_w} \approx  \frac{\sum_{i = 1}^{D} w_i^2d_i}{\sum_{i = 1}^{D} w_id_i}=\hat{\lambda},
\label{eq:varapprox}
\end{equation}
and 
$$d_w=\sum_{i=1}^D w_id_i \sim N\left(Dr_w,  \hat{\lambda}Dr_w(1-r_w) \right).$$
 Then we can repeat the steps above to construct the confidence interval, noting that $\pi_w = f(r_w, p_0, q_0)$. We let the test statistic be
$$\hat{T}_w(\pi)=\frac{\hat{\pi}_{w}-\pi_{w}}{\hat{\sigma}_{w}},$$
where
$\hat{r}_w=d_w/D,$ $\hat{\pi}_w=f(\hat{r}_w, \hat{p}, \hat{q}),$ and
$$\hat{\sigma}^2_w = \frac{\hat{\lambda}\hat{r}_w(1 - \hat{r}_w)}{D(\hat{p} + \hat{q} - 1)^2} + \frac{\hat{\pi}_{w}^2\hat{p}(1 - \hat{p})}{M(\hat{p} + \hat{q} - 1)^2} + \frac{(\hat{r}_{w} - \hat{p})^2\hat{q}(1 - \hat{q})}{N(\hat{p} + \hat{q} - 1)^4},$$
To  calculate the exact $p$-value for testing $H_0: \pi_{w}=\pi$, we only need to modify the step (4) of the algorithm in section 2.1.   
\begin{enumerate}
\item[(4)] For each $(r_k, p_k, q_k)$ from the net, we simulate 
$$ r^*_{wb} \sim N\left(Dr_k, \hat{\lambda}Dr_k(1 - r_k)\right)/D$$
and let
$\pi^*_{wb}=f(r^*_{wb}, p^*_b, q^*_b)$ and
$$\sigma^{*2}_{wb} = \frac{\hat{\lambda}r_{wb}^*(1 - r_{wb}^*)}{D(p_b^* + q_b^* - 1)^2} + \frac{{\pi^*_{wb}}^2p^*_b(1 - p^*_b)}{M(p^*_b + q^*_b - 1)^2} + \frac{(r^*_{wb} - p^*_b)^2q^*_b(1 - q^*_b)}{N(p^*_b + q^*_b - 1)^4}$$
for $b=1, \cdots, B.$ 
\end{enumerate}
In this proposed inference, again we assume that $\lambda_0=\hat{\lambda}$ and the proposed inference is still a hybrid bootstrap method.  Although its coverage may not be "exact" for this reason, we anticipate that it's performance is much more robust than those asymptotic methods. 

\section{Examples}

\begin{figure}
	\centering
	\subfloat[$N$ = 371]{%
		\resizebox*{10cm}{!}{\includegraphics{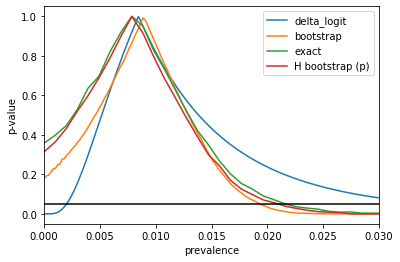}}}\hspace{5pt}
	\subfloat[$N$ = 3324]{%
		\resizebox*{10cm}{!}{\includegraphics{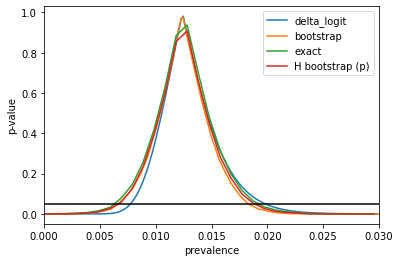}}}
	\caption{Plot of p-values for various values of $\pi$ according to different methods} \label{fig:pvals}
\end{figure}

We first applied our method to analyzing data gathered in \cite{BendavidMS20}. The objective of this study was to estimate the COVID-19 antibody prevalence in Santa Clara county, California, April 2,  2020. The data included $D = 3,330$ volunteers tested for antibody presence. Among them, there were 50 positive test results. Without considering the measurement error, the crude prevalence is 1.5\% with an exact 95\% confidence interval of [1.11, 1.97]\%. To account for the test performance, the reported sensitivity of 130/157 and specificity of 368/371 based on $M = 157$ true positive samples and $N = 371$ true negative samples, respectively, were used to adjust the antibody prevalence estimate in an initial preprint version of the study.  In the meantime, far more extensive additional data on test performance were being collected and verified. We present the corresponding result later. For now,   
$$\hat{r} = \frac{50}{3330}, \hat{p} = \frac{130}{157}, \hat{q} = \frac{368}{371}, D = 3330, M = 157, N = 371.$$
With delta method (with the range preserved logit transformation), the resulting 95\% confidence interval is [0.20, 3.50]\%. The nonparametric bootstrap method yields a similar interval, i.e., [0.00, 1.93]\%.  Then, we applied our methods with $B = 3,000$ and a dense net consisting of 30 evenly spaced points in each $99.9\%$ confidence interval for $r_0, p_0$ and $q_0$ to construct the 95\% exact confidence interval and the hybrid confidence interval fixing $p_0=\hat{p}$. Figure \ref{fig:pvals} plots the estimated exact p-value $\hat{p}(\pi)$, and the corresponding asymptotic p-value based on delta-method, nonparametric bootstrap, and hybrid bootstrap fixing $p_0=\hat{p}.$ It is clearly that $\hat{p}(\pi)$ is higher than its asymptotic counterparts, resulting in a wider confidence interval.
The produced confidence intervals can be found in Table \ref{tab:example1}.  All confidence intervals except that from delta method with logit transformation included 0, and we were unable to make strong conclusions about the lower bound of the prevalence with only a sample size of $N=371$ for estimating the specificity. Since the resulting study cohort may not be randomly sampled from the Santa-Clara population, weighted analysis with individual specific weighting to reflect the demographic makeup of the target population was also conducted. The resulting 95\% confidence interval was [1.18,  3.78]\% and [1.10, 3.69]\% based on delta method and nonparametric bootstrap, respectively. We also constructed the confidence interval based on proposed hybrid bootstrap fixing $\lambda_0=\hat{\lambda}$ and $(\lambda_0, p_0)=(\hat{\lambda}, \hat{p}),$ respectively. The lower ends of the hybrid bootstrap confidence intervals were closer to 0 than that from delta method or nonparametric bootstrap, also suggesting the uncertainty about the lower bound of the prevalence. The basic dilemma was that we cannot reliably differentiate true positives from false positives, since we can't estimate the specificity level with adequate precision. 

To address this difficulty, the study team assembled additional results about the specificity based on 2,953 more measurements, bringing the total number of true negative samples used to estimate the specificity to $N = 3,324$. We had used different meta-analytic methods to combine data across subsets of control samples accounting for potential between-datasets heterogeneity and the results were reasonably ably robust (not shown here). For illustrative purpose, we ignored the potential heterogeneity and assumed simply pooling data in this application was appropriate.  With a larger sample size for estimating specificity ($\hat{q} = 3,308/3,324$, $N = 3,324$), we repeated the construction of 95\% confidence intervals for the unweighted and weighted prevalence (Table \ref{tab:example1}).   These resulting estimates were fairly consistent with those presented in \cite{BendavidMS20}. The unweighted results from different methods were very similar, while the weighted results tended to have modestly wider confidence intervals with the "exact" method and hybrid bootstrap. Figure \ref{fig:pvals} shown that the exact and asymptotic $p$-values were close with each other based on the increased sample size , implying that the distribution of $\hat{\pi}-\pi$ can be approximated well by $N(0, \hat{\sigma}^2)$. The lower bound of the confidence interval based on delta-method and bootstrap for weighted prevalence was slightly higher possibly due to the under-coverage tendency of the bootstrap method at high specificity as our simulation study demonstrated (section 4). 

In order to examine the performance of different methods and the prevailing practices for adjusting for test performances across seroprevalence studies that find low seroprevalence estimates in the tested population, we used a recently published overview of seroprevalence studies \cite{Ioannidis2020}.  In four studies, crude, unadjusted seroprevalence was reported to not exceed 10\% and the authors had tried to adjust for test performance. While three studies from Denmark, the Faroe Islands, and USA \cite{Erikstrup2020, Petersen2020, Havers2020} used the simple bootstrap method to make the statistical inference, the study from Brazil \cite{Hallal2020} implemented a slightly different resampling method. In all four studies, the adjustment for the test performance changed the seroprevalence point estimate by a small amount reflecting the high precision of the test being used.  The analysis results using our proposed methods are summarized in Tables \ref{tab:example2}. The resulting exact 95\% confidence intervals were wider than those based on the delta method and bootstrap. When the number of negative samples used to estimate the specificity was small such as the study from the Faroe Islands, the difference became bigger reflecting the effect of unknown specificity. On the other hand, when the sample sizes used to estimate sensitivity and specificity were adequate and the observed prevalence was not low as in the study at New York, the exact confidence intervals were only slightly wider than those based on simple bootstrap. The data based on which the analysis were conducted can be found in the Appendix in Table \ref{tab:apdataexample3} and some of them were reconstructed from the  results in the published papers \cite{Hallal2020, Havers2020, Erikstrup2020, Petersen2020}.

\begin{table}
	\tbl{The point estimators and 95\% confidence intervals for the weighted and unweighted prevalence in Santa Clara study} 
	{\begin{tabular}{c|c|c|c|c|c}
	\multicolumn{6}{c}{$N=371$}\\
	\hline
			& $\hat{\pi}$\textsuperscript{a} & Delta Method\textsuperscript{b} & Bootstrap & Exact & H Bootstrap $(p_0)$  \\
			\hline
			Unweighted (\%) & 0.85 & (0.20, 3.50) & (0.00, 1.93) & (0.00, 2.06) & (0.00, 2.06)  \\
			Weighted (\%) & 2.80 & (1.18, 3.78)  &(1.10, 3.72)  & (0.29, 5.17) & (0.29, 5.07)\\
\hline
\multicolumn{6}{c}{$N=3,324$}\\  
\hline
			Unweighted (\%) & 1.24 & (0.77, 1.98) & (0.66, 1.84)& (0.68, 1.87) & (0.68, 1.77) \\
			Weighted (\%) & 2.87 & (2.10, 3.6) & (2.12,  3.66) & (1.39, 5.28) & (1.49, 5.08) \\ 
	\end{tabular}}
	\tabnote{\textsuperscript{a}Adjusted for test sensitivity and specificity \textsuperscript{b} Normal Logit Method}
	\label{tab:example1}
\end{table}

\begin{table}
	\tbl{The point estimators and 95\% confidence intervals for the seroprevalence in studies from Brazil, USA, Denmark, and the Faroe Islands.}  
	{\begin{tabular}{c|c|c|c|c|c}
			& $\hat{\pi}$  & Delta Method & Bootstrap & Exact & H Bootstrap $(p_0)$ \\
			\cline{1-6}
			\multicolumn{6}{c}{Brazil}\\
			\hline
			Male (\%) & 0.64 & (0.10, 3.52) & (0.00, 1.55) & (0.00, 1.48) & (0.00, 1.38) \\
			Female (\%) & 0.41 & (0.02, 6.30) & (0.00, 1.30) & (0.00, 1.15) & (0.00, 1.15) \\
	\cline{1-6}
    \multicolumn{6}{c}{USA}\\
    \cline{1-6}
        	\hline
        	Washington Male (\%) & 1.41 & (0.67, 2.95) & (0.33, 2.42) & (0.10, 2.66) & (0.10, 2.56) \\
        	Washington Female (\%) & 1.71 & (0.96, 3.03) & (0.70, 2.64) & (0.39, 2.84) & (0.49, 2.74) \\
        	New York Male (\%) & 5.94 & (4.50, 7.80) & (4.33, 7.59) & (4.17, 7.74) & (4.27, 7.64) \\
        	New York Female (\%) & 5.66 & (4.33, 7.38) & (4.15, 7.23) & (3.98, 7.35) & (4.08, 7.25) \\
	\cline{1-6}
    \multicolumn{6}{c}{Denmark}\\
    \cline{1-6}
			\hline
			Capital (\%) & 3.23 & (2.49, 4.17) & (2.36, 4.06) & (2.13, 4.11) & (2.23, 4.11) \\
			Total (\%) & 1.87 & (1.30, 2.68) & (1.13, 2.48) & (0.78, 2.55) & (0.88, 2.45) \\
	\cline{1-6}
    \multicolumn{6}{c}{Faroe Islands}\\
    \cline{1-6}
			\hline
			Total (\%) & 0.59 & (0.27, 1.31) & (0.19, 1.10) & (0.00, 1.26) & (0.00, 1.16) \\
			Male (\%) & 0.59 & (0.19, 1.82) & (0.00, 1.37) & (0.00, 1.67) & (0.00, 1.57) \\
			Female (\%) & 0.59 & (0.19, 1.82) & (0.00, 1.37) & (0.00, 1.67) & (0.00, 1.57) \\
	\end{tabular}}
	\label{tab:example2}
\end{table}

\section{Simulation}
In this section, we have conducted extensive simulation studies to investigate the operational characteristics of the proposed method and compare it with existing methods in finite sample.

\subsection{Unweighted Inference}

In this simulation study, we  mimic the Santa-Clara study by considering following settings:
\begin{enumerate}
    \item the sensitivity $p_0=83\%;$ sample size $M=157;$
    \item the specificity $q_0 \in \{97\%, 98\%, 98.4\%, 98.6\%, 98.8\%, 99\%, 99.2\%, 99.4\%, 99.8\%, 99.9\%, 100\%\}$; sample size $N=371;$ and separately for $N=3,324;$
    \item the true prevalence $\pi_0\in \{0.4\%, 1.2\%, 5\%, 10\%\}$; sample size $D=3,330.$
\end{enumerate}
For each setting,  we have generated 1,000 sets of data and constructed the 95\% confidence interval using the proposed exact method, the delta method, nonparametric bootstrap, and hybrid bootstraps where we have fixed the proportion of positive tests $(r_0)$, sensitivity ($p_0$), or both. In these simulations, we set $B=500$ for our exact method and nonparametric bootstrap, and $1,000$ for hybrid bootstrap. 
Table \ref{tab:simuresult1} summarizes the average length and the empirical coverage level of constructed 95\% confidence intervals for $\pi_0=1.2\%$ and for $N=317$ only. 

\begin{table}
\tbl{The empirical coverage probability and average length of 95\% confidence interval of $\pi_0$ based on exact method, delta method, nonparametric bootstrap, and hybrid bootstraps; $\pi = 1.2\%, N = 371.$}
{\begin{tabular}{llllccc}
 \hline
 \hline
& Exact Method &Delta Method & Bootstrap & \multicolumn{3}{c}{Hybrid Bootstrap}\\

&              &             &           & $p_0=\hat{p}$ & $r_0=\hat{r}$ & $(p_0, r_0)=(\hat{p}, \hat{r})$ \\
\cline{5-7}
$q_0$ & Cov (Length) & Cov (Length) & Cov (Length) & Cov (Length) & Cov (Length) & Cov (Length) \\ 
   \hline
97.0&0.950 (0.031)&0.934 (0.046) & 0.934 (0.033) & 0.949 (0.030)&0.956 (0.028)&0.947 (0.030)\\
98.0&0.982 (0.028)&0.944 (0.037) & 0.944 (0.028) & 0.957 (0.027)&0.956 (0.024)&0.941 (0.026)\\
98.4&0.970 (0.026)&0.952 (0.033) & 0.948 (0.026) &0.966 (0.025)&0.937 (0.023)&0.943 (0.024)\\
98.6&0.970 (0.025)& 0.920 (0.031) & 0.912 (0.025) &0.961 (0.024)&0.949 (0.022)&0.938 (0.023)\\
98.8&0.974 (0.024)&0.926 (0.029) & 0.914 (0.023) &0.963 (0.023)&0.957 (0.021)&0.941 (0.022)\\
99.0&0.976 (0.023)&0.932 (0.027) & 0.928 (0.022) & 0.952 (0.022)&0.929 (0.020)&0.946 (0.020)\\
99.2&0.970 (0.022)& 0.922 (0.024) & 0.912 (0.021) & 0.961 (0.022)&0.941 (0.019)&0.948 (0.018)\\
99.4&0.980 (0.021)&0.892 (0.021) & 0.902 (0.018) &0.959 (0.020)&0.944 (0.017)&0.963 (0.016)\\
99.6&0.982 (0.019)&0.886 (0.017) & 0.898 (0.016) &0.977 (0.019)&0.969 (0.015)&0.960 (0.014)\\
99.8&0.992 (0.017)&0.908 (0.013) & 0.928 (0.013) &0.993 (0.016)&0.955 (0.013)&0.838 (0.011)\\
99.9&0.992 (0.016)&0.940 (0.011) & 0.950 (0.011) &0.992 (0.015)&0.900 (0.012)&0.673 (0.010)\\
100.0&0.982 (0.015)&0.948 (0.008) & 0.962 (0.008) &0.975 (0.014)&0.768 (0.011)&0.459 (0.009)\\
   \hline
\end{tabular}}
\label{tab:simuresult1}
\end{table}

The exact method always has a coverage of about 95\% or above as anticipated. The delta method and nonparametric bootstrap may result in non-trivial under-coverage when the specificity $q_0$ is high. In addition, two hybrid bootstrap methods that fix $r$ also may produce confidence intervals that are too narrow. One explanation of the failure of these hybrid bootstraps is that when fixing $r_0$ at $\hat{r}$, the sensitivity level implied by the true prevalence $\pi_0$, $$\frac{\hat{r} - (1 - q)(1 - \pi_0)}{\pi_0}>1,$$ 
for all $q$s close to $q_0,$ excluding the true prevalence $\pi_0$ from the confidence interval.
On the other hand, the hybrid bootstrap that only fixes $p_0=\hat{p}$ has coverage at least 95\% for all values of $q_0$. The same pattern repeats in other simulation settings reported in the supplementary materials as well (See Figures \ref{fig:apsinglecov} and \ref{fig:apsinglelen} of the Appendix), so moving forward we focus on the exact method and the hybrid bootstrap that fixes $p_0=\hat{p}.$ 

In the end, we have repeated the simulation with $N=3,324$, the reported sample size in the Santa Clara study after pooling data from multiple sources. The results are summarized in Table 2. In this case where the sample size used to estimate sensitivity is large, these four methods all give reasonable coverage for tested values of $q_0$. Not that in this case, the exact method is not much more conservative than other methods.

\begin{table}
\tbl{The empirical coverage probability and average length of 95\% confidence interval of $\pi_0$ based on exact method, delta method, nonparametric bootstrap, and hybrid bootstrap; $N=3,324.$}
{\begin{tabular}{llllll}
 \hline
 \hline
& & Bootstrap & Delta Method & Exact Method& H Bootstrap \\
& &           &              &             & $p_0=\hat{p}$\\
   \hline

$\pi_0$ & $q_0$ & CovP (Length) & CovP (Length) & CovP (Length) & CovP (Length) \\ 
   \hline
1.2 & 97.0 & 0.954 (0.022) & 0.958 (0.020) & 0.968 (0.021) & 0.951 (0.020) \\
1.2 & 98.0 & 0.958 (0.018) & 0.954 (0.018) & 0.968 (0.018) & 0.953 (0.018) \\
1.2 & 98.4 & 0.952 (0.016) & 0.952 (0.016) & 0.962 (0.017) & 0.953 (0.017) \\
1.2 & 98.6 & 0.960 (0.016) & 0.970 (0.016) & 0.976 (0.017) & 0.959 (0.016) \\
1.2 & 98.8 & 0.956 (0.015) & 0.948 (0.015) & 0.966 (0.016) & 0.956 (0.015) \\
1.2 & 99.0 & 0.954 (0.014) & 0.946 (0.014) & 0.970 (0.015) & 0.969 (0.014) \\
1.2 & 99.2 & 0.964 (0.013) & 0.940 (0.013) & 0.970 (0.014) & 0.948 (0.013) \\
1.2 & 99.4 & 0.936 (0.012) & 0.940 (0.012) & 0.954 (0.013) & 0.967 (0.012) \\
1.2 & 99.6 & 0.954 (0.011) & 0.944 (0.011) & 0.972 (0.012) & 0.953 (0.011) \\
1.2 & 99.8 & 0.962 (0.009) & 0.942 (0.010) & 0.968 (0.010) & 0.957 (0.010) \\
1.2 & 99.9 & 0.944 (0.009) & 0.960 (0.009) & 0.968 (0.009) & 0.957 (0.009) \\
1.2 & 100.0 & 0.942 (0.008) & 0.948 (0.008) & 0.932 (0.008) & 0.913 (0.008) \\
   \hline
\end{tabular}}
\label{tab:simuresult2}
\end{table}
\FloatBarrier

\subsection{Stratum Specific Weighted Inference}
In this case, the sensitivity and specificity are chosen as in section 4.1:
\begin{enumerate}
    \item the sensitivity is 83\%; sample size $M=157;$
    \item the specificity $\in \{97\%, 98\%, 98.4\%, 98.6\%, 98.8\%, 99\%, 99.2\%, 99.4\%, 99.8\%, 99.9\%, 100\%\}$; sample size $N=371$; and separately for $N=3,324.$
\end{enumerate}
The true prevalence is stratum-specific and we have considered six strata summarized in Table \ref{tab:simulationsetting}. The true prevalence for the target population is $\sum_{s=1}^6 w_s\pi_s=1.2\%.$ For comparison purpose, we constructed the 95\% confidence interval using nonparametric bootstrap,  delta method,  proposed hybrid bootstrap fixing $\lambda_0=\hat{\lambda}$, and faster hybrid bootstrap fixing $(\lambda_0, p_0)=(\hat{\lambda}, \hat{p}).$ Table \ref{tab:simuresult3} summarizes the simulation results. The delta-method is slightly better than nonparametric bootstrap but still produces under-covered confidence intervals for some $q_0.$ On the other hand,  the two proposed hybrid bootstrap methods perform satisfactorily. 
\FloatBarrier
\begin{table}
\tbl{Simulation Setting for Stratified Inference}
{\begin{tabular}{lllllll}
 \hline
 \hline
& Strata 1 & Strata 2 & Strata 3 & Strata 4 & Strata 5 & Strata 6 \\
\hline
Weights ($w_s$) &0.05 & 0.07 & 0.08  & 0.15 & 0.25 & 0.40   \\
Prevalence ($\pi_s$) & 0.03\% & 0.70\% & 0.07\% & 0.07\%  & 0.77\% & 2.33\%  \\
Number of tests ($D_s$) & 500 & 700 & 300  &800 & 230 & 800   \\
\hline
\end{tabular}}
\label{tab:simulationsetting}
\end{table}

\begin{table}
\tbl{The empirical coverage probability and average length of 95\% confidence interval of $\pi_0$ based on delta method, nonparametric bootstrap, and hybrid bootstrap for stratum-specific weighted inference}
{\begin{tabular}{llllll}
 \hline
 \hline
& & Bootstrap & Delta Method & H Bootstrap & H Bootstrap \\
& &           &              & $\lambda_0=\hat{\lambda}$ & $(\lambda_0, p_0)=(\hat{\lambda}, \hat{p})$\\
   \hline

$\pi_0$ & $q_0$ & CovP (Length) & CovP (Length) & CovP (Length) & CovP (Length) \\ 
   \hline
1.2 & 97.0 & 0.938 (0.046) & 0.952 (0.034) & 0.964 (0.034) & 0.963 (0.033) \\
1.2 & 98.0 & 0.934 (0.038) & 0.954 (0.030) & 0.962 (0.030) & 0.958 (0.030) \\
1.2 & 98.4 & 0.922 (0.034) & 0.938 (0.028) & 0.964 (0.029) & 0.955 (0.029) \\
1.2 & 98.6 & 0.920 (0.031) & 0.940 (0.027) & 0.968 (0.029) & 0.962 (0.028) \\
1.2 & 98.8 & 0.902 (0.029) & 0.926 (0.025) & 0.962 (0.028) & 0.956 (0.027) \\
1.2 & 99.0 & 0.918 (0.027) & 0.938 (0.024) & 0.978 (0.026) & 0.952 (0.026) \\
1.2 & 99.2 & 0.902 (0.024) & 0.930 (0.022) & 0.972 (0.026) & 0.965 (0.025) \\
1.2 & 99.4 & 0.882 (0.021) & 0.930 (0.020) & 0.970 (0.025) & 0.969 (0.024) \\
1.2 & 99.6 & 0.858 (0.017) & 0.946 (0.018) & 0.986 (0.023) & 0.984 (0.022) \\
1.2 & 99.8 & 0.836 (0.013) & 0.960 (0.015) & 0.996 (0.020) & 0.991 (0.019) \\
1.2 & 99.9 & 0.844 (0.012) & 0.940 (0.014) & 0.984 (0.019) & 0.986 (0.018) \\
1.2 & 100.0 & 0.802 (0.008) & 0.922 (0.011) & 0.970 (0.017) & 0.952 (0.016) \\
   \hline
\end{tabular}}
\label{tab:simuresult3}
\end{table}

\subsection{Individual Specific Weighted Inference}
\begin{figure}
	\centering
	\includegraphics[width=0.66\textwidth]{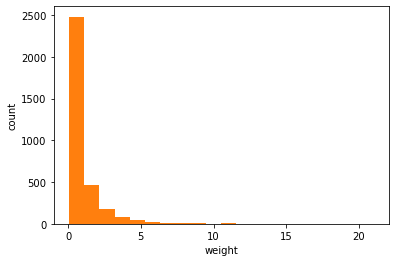}
	\caption{The Distribution of Individual Weights in the Santa Clara Study}
	\label{fig:weights}
\end{figure}

For cases needing individual specific weighting, we adopted the similar simulation settings for sensitivity and specificity in section 4.2. The same individual weights in \cite{BendavidMS20} were used as weights, whose distribution is shown in Figure \ref{fig:weights}. The median weight is 0.48 with an inter-quartile range of [0.22, 1.11].  To specify, $\pi_i,$ the probability of the $i$th individual having the disease or antibody, we let 
$$\pi_i=\frac{\exp(-4.40+0.17w_i)}{1+\exp(-4.40+0.17w_i)}, i=1, \cdots, D=3,330,$$
based on the fitted logistic regression to the observed data in Santa Clara study, where the intercept is adjusted so that the weighted prevalence $D^{-1}\sum_{i=1}^D w_i\pi_i=1.2\%.$ This model suggests  a higher individual-specific weight $w_i$ was associated with a higher probability $\pi_i$. Again, we compared nonparametric bootstrap, delta method, proposed hybrid bootstrap fixing $\lambda_0=\hat{\lambda}$ and faster hybrid bootstrap fixing $(\lambda_0, p_0)=(\hat{\lambda}, \hat{p}).$ The simulation results can be found in Table \ref{tab:simuresult4}. The nonparametric bootstrap performs poorly for most values of $q_0$. The delta method performs reasonably well until for $q_0$ near 1, where the normality breaks down and the coverage starts to decrease drastically. On the other hand, the performance of hybrid bootstrap method fixing $\lambda_0$ is fairly robust in terms of maintaining the appropriate coverage level except when $q_0=100\%.$ The hybrid method fixing both $\lambda_0$ and $p_0$ performs similarly.

\begin{table}
\tbl{The empirical coverage probability and average length of 95\% confidence interval of $\pi_0$ based on delta method, nonparametric bootstrap, and hybrid bootstrap for individual-specific weighted inference}
{\begin{tabular}{llllll}
 \hline
 \hline
& & Bootstrap & Delta Method & H Bootstrap & H Bootstrap \\
       &        &               &               &  $\lambda_0=\hat{\lambda}$             & $(\lambda_0, p_0)=(\hat{\lambda}, \hat{p})$\\
   \hline

$\pi_0$ & $q_0$ & CovP (Length) & CovP (Length) & CovP (Length) & CovP (Length) \\ 
   \hline
1.2 & 97.0 & 0.862 (0.046) & 0.964 (0.043) & 0.974 (0.049) & 0.967 (0.049) \\
1.2 & 98.0 & 0.823 (0.037) & 0.960 (0.038) & 0.968 (0.047) & 0.947 (0.044) \\
1.2 & 98.4 & 0.834 (0.034) & 0.956 (0.036) & 0.958 (0.043) & 0.946 (0.043) \\
1.2 & 98.6 & 0.822 (0.031) & 0.970 (0.034) & 0.942 (0.042) & 0.958 (0.041) \\
1.2 & 98.8 & 0.809 (0.029) & 0.954 (0.033) & 0.962 (0.043) & 0.953 (0.039) \\
1.2 & 99.0 & 0.779 (0.027) & 0.955 (0.032) & 0.954 (0.039) & 0.965 (0.037) \\
1.2 & 99.2 & 0.764 (0.024) & 0.949 (0.030) & 0.968 (0.038) & 0.964 (0.036) \\
1.2 & 99.4 & 0.730 (0.021) & 0.940 (0.028) & 0.968 (0.036) & 0.952 (0.035) \\
1.2 & 99.6 & 0.705 (0.017) & 0.911 (0.026) & 0.960 (0.034) & 0.954 (0.032) \\
1.2 & 99.8 & 0.605 (0.013) & 0.894 (0.024) & 0.938 (0.031) & 0.925 (0.029) \\
1.2 & 99.9 & 0.543 (0.011) & 0.854 (0.022) & 0.926 (0.029) & 0.889 (0.027) \\
1.2 & 100.0 & 0.499 (0.008) & 0.780 (0.020) & 0.860 (0.027) & 0.864 (0.026) \\
   \hline
\end{tabular}}
\label{tab:simuresult4}
\end{table}

\section{Discussion}
In order to estimate the prevalence of a disease using imperfect tests, we developed a method that provides confidence intervals with appropriate coverage. This is important because in many scenarios there is not enough data for large sample approximations to be accurate, especially when the sensitivity $p_0$ or specificity $q_0$ is very close to 1, which can cause the naive bootstrap confidence intervals to be too narrow. However, our method is computationally more expensive than the bootstrap method by several orders of magnitude, which translates to about half a minute to compute a single confidence interval on a PC with a Ryzen 3900X CPU. In practice, we don't believe this will impose too large a burden, as typically there is no need to compute a confidence interval many times. 

In addition, only the proposed method for unweighted inference is truly exact; in two weighted cases we still make some approximations for the distribution of $r_w$. Such an approximation is unavoidable due to the fact that the variance inflation factor $\lambda_0$ is unknown and may not be estimated well empirically. Also, we note that the performance of the simple bootstrap becomes better as the sample size for estimating specificity $N$ rises. Therefore, while the sample size for estimating prevalence $D$ is important, the size of the confidence interval also heavily depends on the sample size for estimating sensitivity and specificity, and especially the latter. Even as $D$ grows, the length of the confidence interval will not shrink to zero, since the uncertainty of the sensitivity and specificity affects the estimation of the true prevalence. For experiments aiming to estimate prevalence in settings where low values are expected, it is worth the effort to accurately estimate the sensitivity and specificity. This prerequisite is no longer a serious issue when the prevalence is sufficiently high.

Our review of the literature of COVID-19 seroprevalence studies \cite{deSouzaBC20, Ceylan20, SignorelliSO20, HuSD20} shows that many studies that estimate low crude prevalence do not even try to adjust for test performance. Some of them may try to validate the positive samples using a different laboratory assay \cite{Ng2020}. Many others may assume that specificity is perfect. For well-validated assays, this assumption may be approximately correct. For example, in the case of the assay used in the Santa Clara study, the specificity was 99.5-99.8\% depending on how pooling or meta-analysis of control datasets would be performed. Moreover, among the few control samples coined as ``false positives'', the majority were probably true positives that had been mischaracterized, as these control samples came from data collected during the COVID-19 pandemic, where a negative RT-PCR result can not rule out the possibility that a person had already been infected in the past. Most of the remaining ``false positives'' that came from pre-COVID samples were atypical cases (e.g. from people with extremely high titers of rheumatoid factor) that are rarely encountered in the general population. This means the true specificity of the test used in Santa Clara study may be even higher.  However, our simulation study shows that the simple bootstrap or delta-method may still yield suboptimal coverage even with a perfect specificity and  the method that we propose may have value in such a setting. 

Another strategy to alleviate the false positive issue would be via study design: to retest all patients whose results are positive \cite{sempos20}. An important reason why it is difficult to estimate the prevalence is because the false positive rate can be relatively high, and the estimated prevalence is very sensitive to the false positive rate. If it is possible to have an independent second test applied to patients who are test positive, we can eliminate almost all false positives and overcome this obstacle.

\bibliographystyle{tfs}
\bibliography{sections/biblio}

\newcommand{\noopsort}[1]{} \newcommand{\printfirst}[2]{#1}
  \newcommand{\singleletter}[1]{#1} \newcommand{\switchargs}[2]{#2#1}
\begin{thebibliography}{10}
\providecommand{\MR}{\relax\unskip\space MR }
\providecommand{\url}[1]{\normalfont{#1}}
\providecommand{\urlprefix}{Available at }

\bibitem{BendavidMS20}
E. Bendavid, B. Mulaney, N. Sood, S. Shah, E. Ling, R. Bromley-Dulfano, C. Lai,
  Z. Weissberg, R. Saavedra-Walker, J. Tedrow, D. Tversky, A. Bogan, T. Kupiec,
  D. Eichner, R. Gupta, J. Ioannidis, and J. Bhattacharya, \emph{{COVID}-19
  antibody seroprevalence in santa clara county, california}  (2020).
  \urlprefix\url{https://doi.org/10.1101/2020.04.14.20062463}.

\bibitem{Ceylan20}
Z. Ceylan, \emph{Estimation of {COVID}-19 prevalence in italy, spain, and
  france}, Science of The Total Environment 729 (2020), p. 138817.

\bibitem{ChanZ99}
I.S.F. Chan and Z. Zhang, \emph{Test-based exact confidence intervals for the
  difference of two binomial proportions}, Biometrics 55 (1999), pp.
  1202--1209.

\bibitem{ChuangL00}
C.S. Chuang and T.L. Lai, \emph{Hybrid resampling methods for confidence
  intervals}, Statistica Sinica 10 (2000), pp. 1--33.

\bibitem{deSouzaBC20}
W. de  Souza, L. Buss, D. Candido, and  et  al., \emph{Epidemiological and
  clinical characteristics of the {COVID}-19 epidemic in brazil}, Nature Human
  Behaviour 4 (2020), pp. 856--865.

\bibitem{EnoeGJ00}
C. Enøe, M.P. Georgiadis, and W.O. Johnson, \emph{Estimation of sensitivity
  and specificity of diagnostic tests and disease prevalence when the true
  disease state is unknown}, Preventive Veterinary Medicine 45 (2000), pp. 61
  -- 81.

\bibitem{Erikstrup2020}
C. Erikstrup, C.E. Hother, O.B.V. Pedersen, K. M{\o}lbak, R.L. Skov, D.K. Holm,
  S. S{\ae}kmose, A.C. Nilsson, P.T. Brooks, J.K. Boldsen, C. Mikkelsen, M.
  Gybel-Brask, E. S{\o}rensen, K.M. Dinh, S. Mikkelsen, B.K. M{\o}ller, T.
  Haunstrup, L. Harritsh{\o}j, B.A. Jensen, H. Hjalgrim, S.T. Lillevang, and H.
  Ullum, \emph{Estimation of {SARS}-{CoV}-2 infection fatality rate by
  real-time antibody screening of blood donors}  (2020).
  \urlprefix\url{https://doi.org/10.1101/2020.04.24.20075291}.

\bibitem{FeldmanC98}
G.J. Feldman and R.D. Cousins, \emph{Unified approach to the classical
  statistical analysis of small signals}, Phys. Rev. D 57 (1998), pp.
  3873--3889.

\bibitem{gronsbell2020exact}
J. Gronsbell, C. Hong, L. Nie, Y. Lu, and L. Tian, \emph{Exact inference for
  the random-effect model for meta-analyses with rare events}, Statistics in
  Medicine 39 (2020), pp. 252--264.

\bibitem{Hallal2020}
P.C. Hallal, F.P. Hartwig, B.L. Horta, M.F. Silveira, C.J. Struchiner, L.P.
  Vidaletti, N.A. Neumann, L.C. Pellanda, O.A. Dellagostin, M.N. Burattini,
  G.D. Victora, A.M.B. Menezes, F.C. Barros, A.J.D. Barros, and C.G. Victora,
  \emph{{SARS}-{CoV}-2 antibody prevalence in brazil: results from two
  successive nationwide serological household surveys}, The Lancet Global
  Health 8 (2020), pp. e1390--e1398.

\bibitem{Havers2020}
F.P. Havers, C. Reed, T. Lim, J.M. Montgomery, J.D. Klena, A.J. Hall, A.M. Fry,
  D.L. Cannon, C.F. Chiang, A. Gibbons, I. Krapiunaya, M. Morales-Betoulle, K.
  Roguski, M.A.U. Rasheed, B. Freeman, S. Lester, L. Mills, D.S. Carroll, S.M.
  Owen, J.A. Johnson, V. Semenova, C. Blackmore, D. Blog, S.J. Chai, A. Dunn,
  J. Hand, S. Jain, S. Lindquist, R. Lynfield, S. Pritchard, T. Sokol, L. Sosa,
  G. Turabelidze, S.M. Watkins, J. Wiesman, R.W. Williams, S. Yendell, J.
  Schiffer, and N.J. Thornburg, \emph{Seroprevalence of antibodies to
  {SARS}-{CoV}-2 in 10 sites in the united states, march 23-may 12, 2020},
  {JAMA} Internal Medicine  (2020).
  \urlprefix\url{https://doi.org/10.1001/jamainternmed.2020.4130}.

\bibitem{HuSD20}
Y. Hu, J. Sun, Z. Dai, H. Deng, X. Li, Q. Huang, Y. Wu, L. Sun, and Y. Xu,
  \emph{Prevalence and severity of corona virus disease 2019 ({COVID}-19): A
  systematic review and meta-analysis}, Journal of Clinical Virology 127
  (2020), p. 104371.

\bibitem{Ioannidis2020}
J. Ioannidis, \emph{The infection fatality rate of {COVID}-19 inferred from
  seroprevalence data}  (2020).
  \urlprefix\url{https://doi.org/10.1101/2020.05.13.20101253}.

\bibitem{michael2019exact}
H. Michael, S. Thornton, M. Xie, and L. Tian, \emph{Exact inference on the
  random-effects model for meta-analyses with few studies}, Biometrics 75
  (2019), pp. 485--493.

\bibitem{Ng2020}
D.L. Ng, G.M. Goldgof, B.R. Shy, A.G. Levine, J. Balcerek, S.P. Bapat, J.
  Prostko, M. Rodgers, K. Coller, S. Pearce, S. Franz, L. Du, M. Stone, S.K.
  Pillai, A. Sotomayor-Gonzalez, V. Servellita, C.S.S. Martin, A. Granados,
  D.R. Glasner, L.M. Han, K. Truong, N. Akagi, D.N. Nguyen, N.M. Neumann, D.
  Qazi, E. Hsu, W. Gu, Y.A. Santos, B. Custer, V. Green, P. Williamson, N.K.
  Hills, C.M. Lu, J.D. Whitman, S.L. Stramer, C. Wang, K. Reyes, J.M.C. Hakim,
  K. Sujishi, F. Alazzeh, L. Pham, E. Thornborrow, C.Y. Oon, S. Miller, T.
  Kurtz, G. Simmons, J. Hackett, M.P. Busch, and C.Y. Chiu,
  \emph{{SARS}-{CoV}-2 seroprevalence and neutralizing activity in donor and
  patient blood}, Nature Communications 11 (2020).
  \urlprefix\url{https://doi.org/10.1038/s41467-020-18468-8}.

\bibitem{Petersen2020}
M.S. Petersen, M. Str{\o}m, D.H. Christiansen, J.P. Fjallsbak, E.H. Eliasen, M.
  Johansen, A.S. Veyhe, M.F. Kristiansen, S. Gaini, L.F. M{\o}ller, B. Steig,
  and P. Weihe, \emph{Seroprevalence of {SARS}-{CoV}-2{\textendash}specific
  antibodies, faroe islands}, Emerging Infectious Diseases 26 (2020), pp.
  2760--2762.

\bibitem{ReiczigelFO10}
J. Reiczigel, J. Foldi, and L. Ózsvari, \emph{Exact confidence limits for
  prevalence of a disease with an imperfect diagnostic test}, Epidemiology and
  Infection 138 (2010), p. 1674–1678.

\bibitem{sempos20}
C. Sempos and L. Tian, \emph{Adjusting coronavirus prevalence estimates for
  laboratory test kid error}, American Journal of Epidemiology (in press)
  (2020).

\bibitem{SenWW09}
B. Sen, M. Walker, and M. Woodroofe, \emph{On the unified method with nuisance
  parameters}, Statistica Sinica 19 (2009), pp. 301--314.

\bibitem{SignorelliSO20}
C. Signorelli, T. Scognamiglio, and A. Odone, \emph{{COVID}-19 in italy: impact
  of containment measures and prevalence estimates of infection in the general
  population}, Acta Bio Medica Atenei Parmensis 91 (2020), pp. 175--179.

\end{thebibliography}

\newpage
\section{Appendix}

\subsection{Additional Simulation Results and Data used for the seroprevalence in studies from Brazil, USA, Denmark, and the Faroe Islands}

\begin{figure}[ht]
	\centering
	\subfloat{%
		\resizebox*{9cm}{!}{\includegraphics{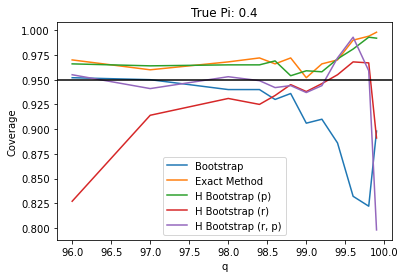}}}\hspace{5pt}
	\subfloat{%
		\resizebox*{9cm}{!}{\includegraphics{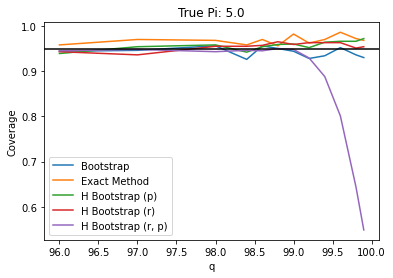}}}\hspace{5pt}
	\subfloat{%
		\resizebox*{9cm}{!}{\includegraphics{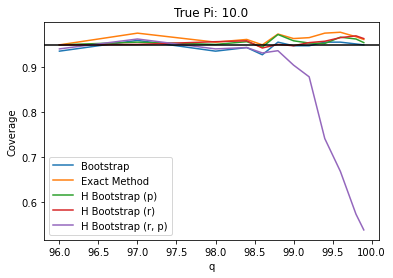}}}
	\caption{Plot of coverages for varying values of specificity $q$ for $N = 371$ under $\pi = 0.4, 5.0, 10.0$} \label{fig:apsinglecov}
\end{figure}

\begin{figure}[ht]
	\centering
	\subfloat{%
		\resizebox*{9cm}{!}{\includegraphics{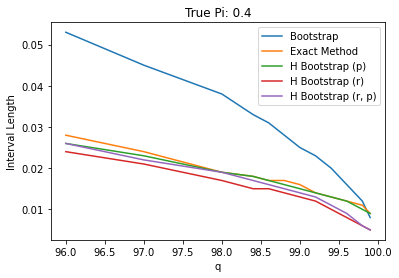}}}\hspace{5pt}
	\subfloat{%
		\resizebox*{9cm}{!}{\includegraphics{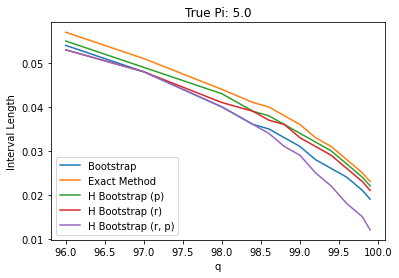}}}\hspace{5pt}
	\subfloat{%
		\resizebox*{9cm}{!}{\includegraphics{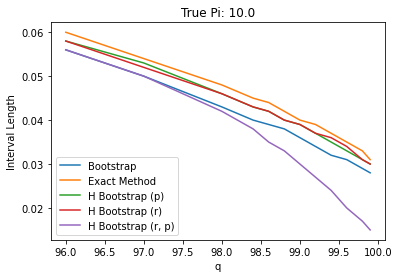}}}
	\caption{Plot of confidence interval lengths for varying values of specificity $q$ for $N = 371$ under $\pi = 0.4, 5.0, 10.0$} \label{fig:apsinglelen}
\end{figure}

In Figure \ref{fig:apsinglecov}, we plot the empirical coverage levels of various confidence intervals assuming different true prevalence level, i.e., $\pi_0\in \{0.4\%, 5\%, 10\%\}.$ While most confidence intervals retain appropriate coverage level when $\pi_0=10\%,$  only the proposed exact method and hybrid bootstrap fixing $p_0$ at $\hat{p}$ perform satisfactorily when the prevalence $\pi_0=0.4\%.$ Specifically, even when the prevalence is 5\%, the 95\% confidence interval based on nonparametric bootstrap may still too liberal with a coverage level approximately 90\% for some specificity values.  Figure \ref{fig:apsinglelen} plots the average length of the 95\% confidence intervals. Note that the average length of the proposal exact confidence interval is not substantially longer than alternatives. 

Table \ref{tab:apdataexample3} includes the data used for the analysis of the seroprevalence in studies from Brazil, USA, Denmark and the Faroe Islands. Note that some studies only reported the confidence intervals for the test sensitivity and specificity and the corresponding data were reconstructed based on the confidence interval, which may be slightly different from the actual data.

\FloatBarrier
\begin{table}[ht]
	\tbl{Data used for the seroprevalence in studies from Brazil, USA, Denmark, and the Faroe Islands.} 
	{\begin{tabular}{c|c|c|c|c|c|c}
			& $r$ (\%) & $p$(\%) & $q$(\%) & $d/D$ & $m/M$ & $n/N$  \\
			\cline{1-7}
			\multicolumn{7}{c}{Brazil}\\
			\hline
			Male & 1.50 & 84.79 & 99.03 & 158/10531 & 446/526 & 513/518 \\
			Female & 1.31 & 84.79 & 99.03 & 189/14464 & 446/526 & 513/518 \\
	\cline{1-7}
    \multicolumn{7}{c}{USA}\\
    \cline{1-7}
        	\hline
        	Washington Male & 1.95 & 96.00 & 99.40 & 26/1334 & 96/100 & 497/500 \\
        	Washington Female & 2.23 & 96.00 & 99.40 & 43/1930 & 96/100 & 497/500 \\
        	New York Male & 6.27 & 96.00 & 99.40 & 72/1149 & 96/100 & 497/500 \\
        	New York Female & 6.00 & 96.00 & 99.40 & 80/1333 & 96/100 & 497/500 \\
	\cline{1-7}
    \multicolumn{7}{c}{Denmark}\\
    \cline{1-7}
			\hline
			Capital & 3.11 & 82.58 & 99.54 & 203/6528 & 128/155 & 648/651 \\
			Total & 2.00 & 82.58 & 99.54 & 412/20640 & 128/155 & 648/651 \\
	\cline{1-7}
    \multicolumn{7}{c}{Faroe Islands}\\
    \cline{1-7}
			\hline
			Total & 0.56 & 94.44 & 100.00 & 6/1075 & 238/252 & 308/308 \\
			Male & 0.56 & 94.44 & 100.00 & 3/538 & 238/252 & 308/308 \\
			Female & 0.56 & 94.44 & 100.00 & 3/537 & 238/252 & 308/308 \\
	\end{tabular}}
	\label{tab:apdataexample3}
\end{table}

\end{document}